\documentclass[a4paper,11pt]{article}
\usepackage{pos}

\title{Multi-messenger constraints on transient accelerators of ultra-high energy cosmic rays}

\author*[a]{Antonio Condorelli}
\author[b,c]{Jonathan Biteau}
\author[b]{Olivier Deligny}
\author[d]{Remi Adam}

\affiliation[a]{Universit{\'e} Paris Cit{\'e}, CNRS, Astroparticule et Cosmologie, F-75013 Paris, France}

\affiliation[b]{Université Paris-Saclay, CNRS/IN2P3, IJCLab,
15 Rue Georges Clemenceau, 91405 Orsay, France}
\affiliation[c]{Institut universitaire de France (IUF)}

\affiliation[d]{Université Côte d'Azur, Observatoire de la Côte d'Azur, CNRS, Laboratoire Lagrange, France}

\emailAdd{condorelli@apc.in2p3.fr}

\abstract{The origin of ultra-high-energy cosmic rays (UHECRs) remains an open questions in astrophysics. We explore two primary scenarios for the distribution of UHECR sources, assuming that their production rate follows either the cosmic star-formation-rate or stellar-mass density. By jointly fitting the UHECR energy spectrum and mass composition measured by the Pierre Auger Observatory above the ankle ($10^{18.7}$ eV), we derive constraints on the acceleration mechanisms, source energetics, and elemental abundances at escape. Using these constraints, we generate sky maps above 40 EeV based on a catalog of over 400,000 galaxies out to 350 Mpc, providing a near-infrared flux-limited sample that maps the two stellar-activity tracers across the full sky.
A crucial factor in understanding UHECR propagation is the influence of large-scale cosmic structures, particularly galaxy clusters—the largest gravitationally bound systems in the Universe, which are filled with magnetized diffuse plasma. Intermittent sources hosted in galaxies within such structures, coupled with cosmic magnetic fields, shape the observed UHECR arrival directions and provide insights into the burst rate of the sources. We show that these environments can significantly impact UHECR transport, making them particularly opaque to heavy nuclei. Additionally, we compute the expected secondary neutrino and photon fluxes from UHECR interactions in these environments and compare them with current experimental limits, constraining the maximum energy that particles can achieve. Finally, we assess the compatibility of these constraints with astrophysical candidates, identifying long gamma-ray bursts as the most promising sources.}

\ConferenceLogo{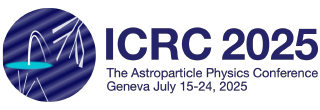}

\FullConference{39th International Cosmic Ray Conference (ICRC2025)\\
 15–24 July 2025\\
Geneva, Switzerland\\}

\begin{document}
\maketitle

\section{Introduction} 
\label{sec:intro}

The sources of charged cosmic rays with energies exceeding 1 EeV remain unknown. Nonetheless, observations of the energy spectrum and mass composition of ultra-high energy cosmic rays (UHECRs) by the Pierre Auger Observatory over the past two decades have significantly constrained the characteristics of the acceleration mechanisms, source energetics, and nuclear composition at the sources~\citep{PierreAuger:2021hun,PierreAuger:2014sui}. If electromagnetic processes dominate the acceleration, nuclei with atomic mass $A$ are expected to reach a maximum energy $E^{\mathrm{max}}{Z_A}$ proportional to their electric charge $Z_A$, resulting in a rigidity-dependent cutoff in the intensity of each nuclear species~\citep{PierreAuger:2016use}. This interpretation is supported by the observed increase in the average logarithmic mass, $\langle \ln A \rangle$, with energy. The measured composition points to a dominance of intermediate-mass nuclei, from helium to silicon, accelerated up to $E^{\mathrm{max}}{Z_A} \propto 5 Z_A$ EeV, and escaping the sources with a hard spectral index $\gamma$ (defined via $dN/dE \propto E^{-\gamma}$).

To account for the observed energy density of UHECRs, the required production rate density is approximately $\mathcal{L} \simeq 10^{45}$erg Mpc$^{-3}$ yr$^{-1}$. Assuming that the UHECR luminosity is comparable to the electromagnetic luminosity of candidate sources, this energy budget can be used to narrow down viable source classes~\cite{Alves_Batista_2019}. These candidates range from frequent, low-luminosity sources to rare, high-luminosity transients.

Sky maps of UHECR arrival directions support their extragalactic origin above the ankle (5 EeV), and may carry further information capable of breaking degeneracies between source scenarios. In particular, if UHECRs originate from transient events, their arrival patterns are affected by magnetic deflections, which delay their propagation in a rigidity-dependent way. As a result, the number of visible sources decreases with energy more rapidly than in steady-state scenarios, and different transient rates yield distinct anisotropy signatures.

The aim of this work is to extend previous studies on UHECR source rate density by incorporating multi-messenger constraints from galaxy clusters. Building on the framework established in~\cite{Marafico:2024}, we investigate the hypothesis that clusters of galaxies host transient periods of UHECR production. By combining the observed UHECR arrival directions with complementary astrophysical probes, such as X-ray, gamma-ray, and neutrino emissions, we seek to refine constraints on the rate and distribution of potential UHECR accelerators embedded in these large-scale structures.

\section{Model Description}
\label{sec:model}

This study builds upon the modeling framework developed in~\cite{Marafico:2024}, where a near-infrared flux-limited galaxy catalog~\cite{Biteau:2021pru} was used to reconstruct the 3D distribution of galaxies within 350~Mpc. This catalog, containing about 400,000 galaxies with stellar mass ($M_\star$) and star formation rate (SFR) estimates, incorporates spectro-photometric distances, corrections for incompleteness, and large-scale structure features such as the Local Sheet~\cite{McCall:2014eha} and Cosmic Flows dark matter fields~\cite{Hoffman:2017ako}. Beyond 350~Mpc, the density field was extrapolated using the model from~\cite{Koushan:2021qms}.

In the model, each galaxy hosted transient UHECR sources bursting at a rate proportional to either $M_\star$ or SFR. The emission spectrum was modeled as an exponentially suppressed power law, with parameters for protons and heavier nuclei as in~\cite{Luce:2022awd}. Contributions to the observed flux included both resolved galaxies within 350~Mpc and an isotropic background beyond that distance. Propagation was handled with SimProp v2r4~\citep{Aloisio:2017iyh}.

Magnetic deflections were modeled using a von Mises–Fisher distribution with a concentration parameter encoding the cumulative effect of Galactic, local, and extragalactic magnetic fields. For transient sources, the observable duration $\Delta\tau$ due to magnetic dispersion was used to compute the expected number of past bursts ($\lambda_i$), assuming Poisson statistics:
\begin{equation}
    \lambda_i = \dot{k} M_{\star i} \Delta\tau,
\end{equation}
with $\dot{k}$ the burst rate per unit stellar mass. 

\section{The importance of Galaxy clusters}
\begin{figure}[h!]
\begin{center}
\begin{minipage}{0.49\textwidth}
\centering
  \includegraphics[scale=0.45]{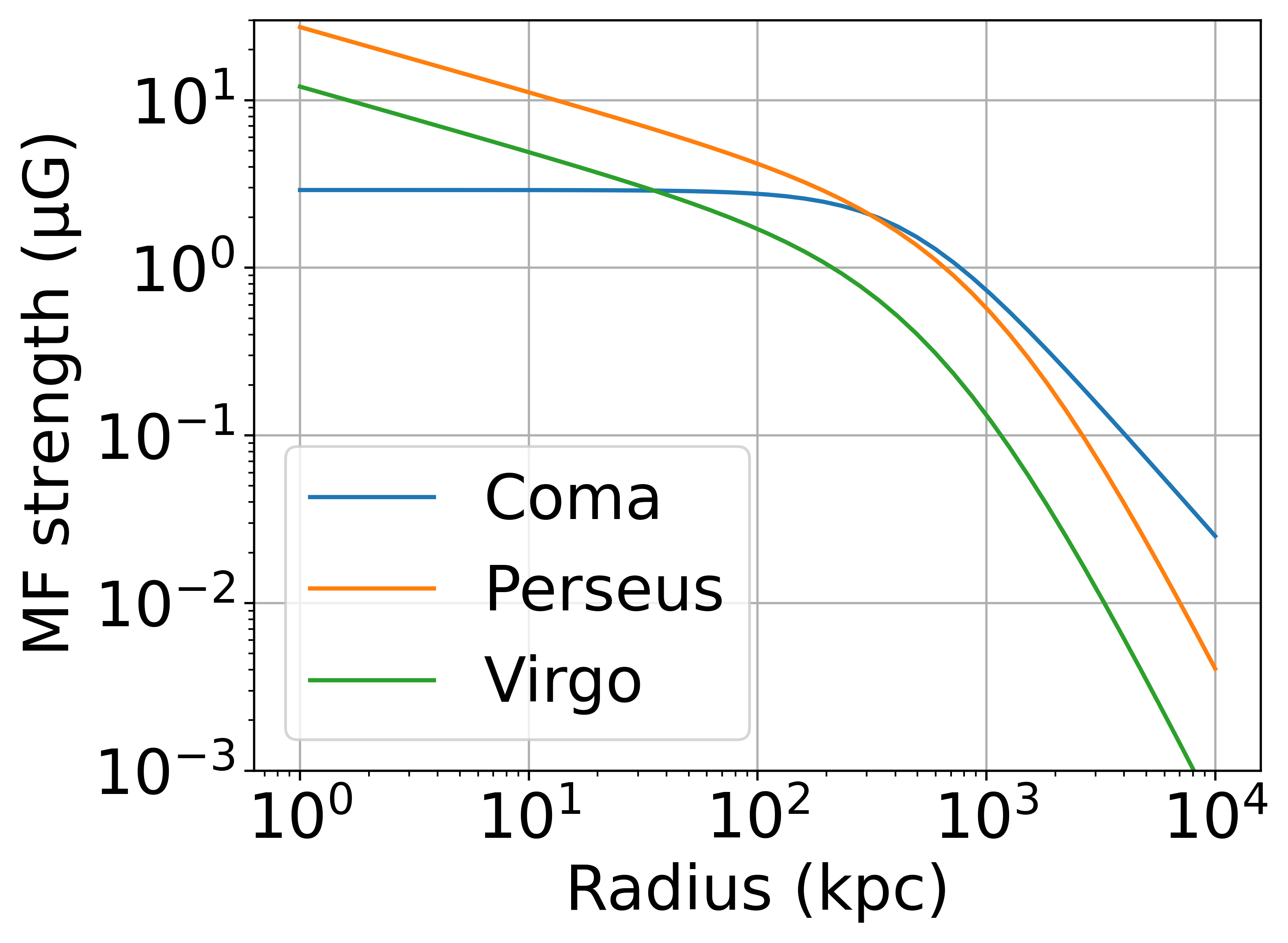}
        \end{minipage}
        \begin{minipage}{0.49\textwidth}
\centering
        \includegraphics[scale=.45]{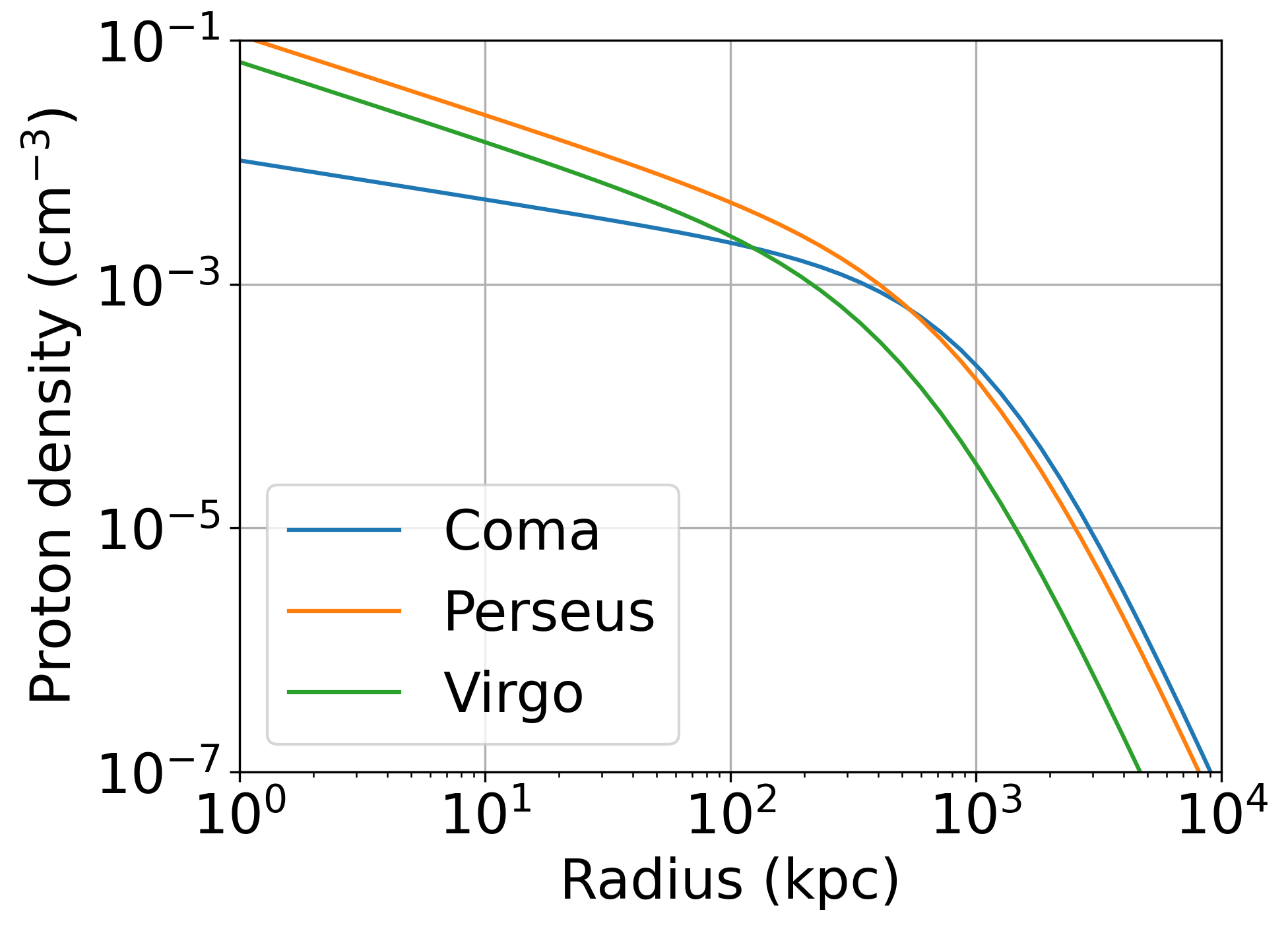}
        \end{minipage}
  \caption{Left panel: Magnetic field profiles adopted for Virgo, Perseus and Coma. Right panel: Gas density profiles adopted for Virgo, Perseus and Coma. In both scenarios, the assumptions underlying these expectations are described in detail in~\cite{Condorelli}. }
  \label{fig:cluster_prof}
\end{center}
\end{figure}

In \cite{Condorelli}, the intracluster medium (ICM) was modeled under a self-similar assumption, allowing the radial pressure profile of each cluster to be described using only its mass and redshift. The magnetic field is assumed to scale with the thermal energy density of the ICM, with the normalization calibrated on the Perseus cluster~[$(l, b) = (150.6^\circ, -13.3^\circ)$, $d \simeq 70$~Mpc, $M \simeq 5.8 \times 10^{14}\, M_\odot$, $R_{500} \simeq 1.3$~Mpc]. Based on these ingredients, the propagation of UHECRs through the gas-density and magnetic-field profiles can be simulated for any given cluster. Among clusters within the GZK horizon, Virgo is the brightest in X-rays~[$(l, b) = (279.7^\circ, 74.5^\circ)$, $d \simeq 15$~Mpc, $M \simeq 1.2 \times 10^{14}\, M_\odot$, $R_{500} \simeq 0.8$~Mpc], followed by Perseus, which is $1.5$--$2$ times fainter. Both clusters are effectively opaque to UHECR protons and nuclei in the rigidity range of interest. In our modeling, the transparency functions from~\cite{Condorelli} are applied to all galaxies located behind disks of radius $3R_{500}$ centered on Virgo and Perseus. The Coma cluster, roughly five times fainter in X-rays than Virgo, is considered subdominant in this context.
The magnetic field and gas density profiles of the three clusters taken into account are shown in figure \ref{fig:cluster_prof}.

To constrain transient sources of UHECRs, the impact of galaxy clusters on cosmic-ray propagation has been investigated~\cite{Condorelli}. In particular, transparency functions that account for magnetic confinement and the suppression of escaping fluxes from dense environments were introduced. This correction is especially relevant for nearby clusters such as Virgo ($d \simeq 15$~Mpc), from which no significant excess is observed in current UHECR sky maps~\cite{PierreAuger:2023mvf}. Accounting for these effects is essential to accurately reproduce the observed angular distribution and energy spectrum of UHECRs at Earth. The propagation of UHECRs through galaxy clusters naturally leads to hadronic and photohadronic interactions with the dense intracluster medium and ambient radiation fields.  These processes inevitably produce secondary neutrinos and gamma rays, which we incorporate directly into our SimProp-based simulations by tracking meson production, decay kinematics, and electromagnetic cascading within each cluster environment. Additionally, we compute the resulting diffuse neutrino and photon fluxes emerging from all clusters in our simulations, in order to place stringent constraints on the rate and on the emissivity by transient cosmic accelerators residing in these dense, magnetized environments.

\section{Multi-messenger signature in galaxy clusters}
\begin{figure}[htp]
\centering
\includegraphics[width=0.64\columnwidth]{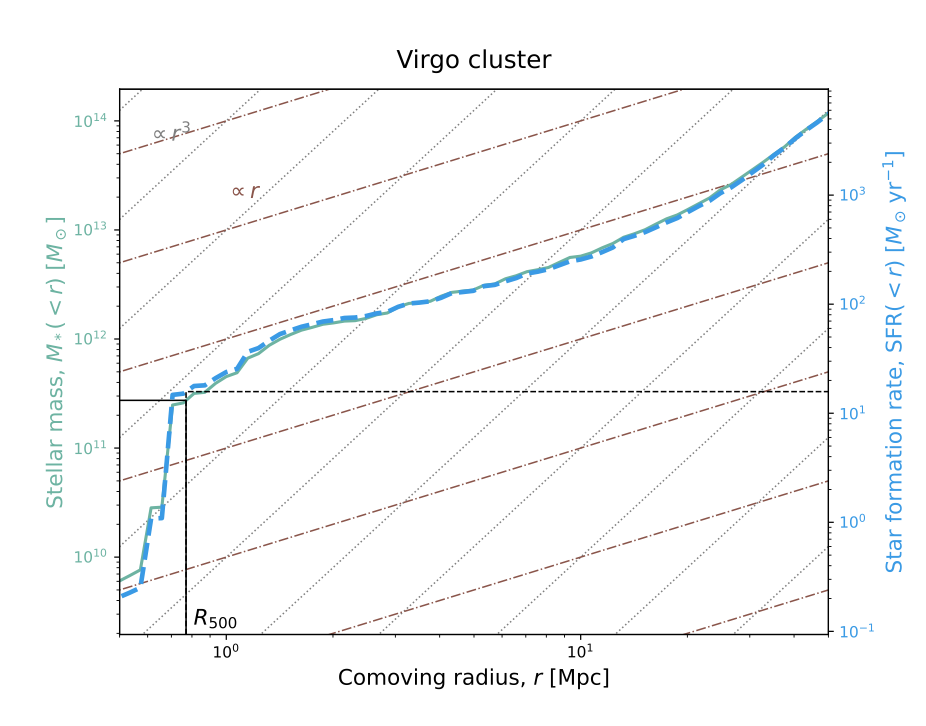}
\caption{Cumulative stellar mass (in solar masses) and star formation density as a function of the comoving radius for Virgo cluster. $R_{500}$ is the radius within which the cluster density is 500 times the critical density of the universe at the cluster redshift, and it is a crucial parameter to describe clusters environment.}
\label{fig:virgo_rate}
\end{figure}
\begin{figure}[htp]
\centering
\includegraphics[width=0.64\columnwidth]{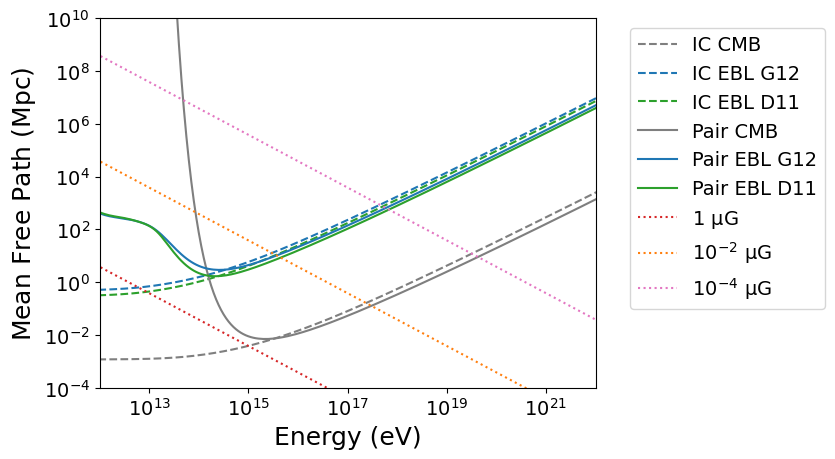}
\caption{Relevant timescale in galaxy clusters. Inverse compton (IC) on CMB and different EBL models are shown with dashed lines; pair production timescales are shown with solid line, while synchrotron timescales are shown in dotted lines for different magnetic field values.}
\label{fig:timescale}
\end{figure}

Figure~\ref{fig:virgo_rate} shows the cumulative stellar mass and star–formation rate profiles for the Virgo cluster:
Over 90\% of the cluster stellar mass—and hence its UHECR burst potential—is contained within a few $R_{500}$.  Since the star–formation rate follows the similar radial dependence, most secondary neutrinos and photons from UHECR interactions will originate within the core region.
From the stellar mass inside, considering the best fit parameters obtained in ~\cite{Marafico:2024}, we can estimate the secondary photon fluxes emitted in the source environment. Since the environment is dominated by hadronic interactions, we expect that the two secondary fluxes are correlated.
In hadronic interactions (e.g.\ $pp$ or $p\gamma$), roughly (within a factor 2) equal numbers of neutral and charged pions are produced.  The neutral pions decay promptly via 
\[
\pi^0 \to \gamma + \gamma,
\]
while each charged pion decays via 
\[
\pi^\pm \to \mu^\pm + \nu_\mu \quad (\to e^\pm + \nu_e + \bar\nu_\mu),
\]
so that the total energy channeled into neutrinos is comparable to that in gamma rays.  Under the assumption of approximate energy equipartition among the $\pi^0,\pi^+,\pi^-$ channels, one therefore expects the secondary neutrino and gamma‑ray fluxes to have similar normalizations and spectral shapes.  \\
Despite the fact the two originate from (almost) the same channels, they face a different fate. In fact, neutrino flux, once produced, is directly propagated to Earth, while, for photons, absorption in cluster environment and in the extragalactic medium are fundamental.\\
To model high-energy particle propagation in the Virgo Cluster, we utilized \texttt{CRPropa} \cite{AlvesBatista2016} to simulate a three-dimensional environment incorporating relevant astrophysical interactions and photon fields. The cluster was approximated as a spherical region of radius $3 \times R_500$, with a radially dependent magnetic field satisfying $\nabla \cdot \mathbf{B} = 0$. Key energy loss processes, including pair production, inverse Compton scattering, and synchrotron radiation, were included. The photon backgrounds considered were the cosmic microwave background (CMB) and extragalactic background light (EBL), while contributions from intra-cluster light and the cosmic radio background were confirmed to be negligible. The timescales relevant in galaxy clusters are shown in figure \ref{fig:timescale}.

\begin{figure}[h!]
\begin{center}
\begin{minipage}{0.49\textwidth}
\centering
  \includegraphics[scale=0.8]{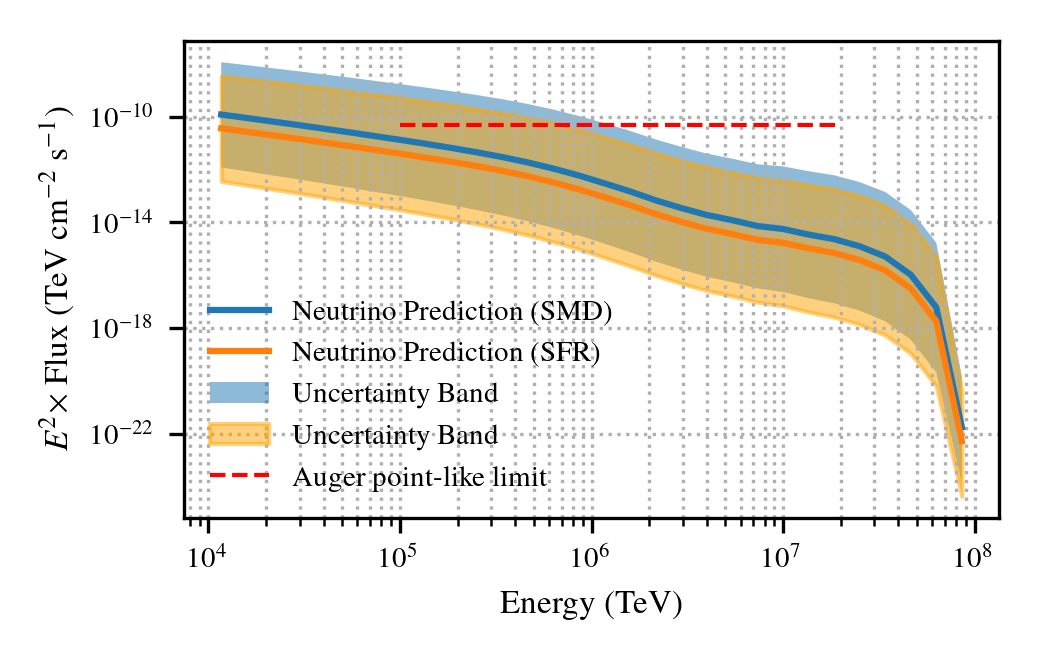}
        \end{minipage}
        \begin{minipage}{0.49\textwidth}
\centering
        \includegraphics[scale=.35]{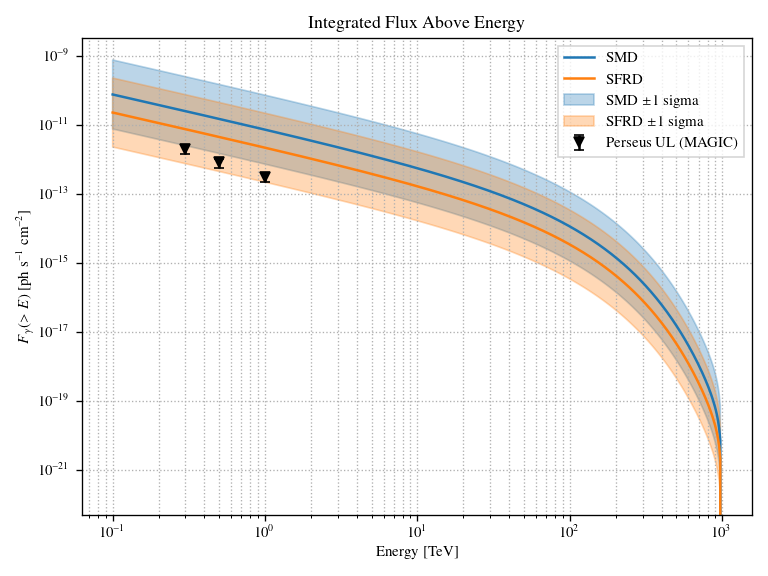}
        \end{minipage}
  \caption{Expected neutrino flux (left) and photon flux (right) at Earth for SMD and SFR scenarios. For the neutrino flux, only corrections due to the adiabatic expansion of the Universe are accounted for, in comparison with Auger limit for point-like sources \cite{Auger2019PointNeutrino}. In contrast, the photon flux includes attenuation due to propagation through the intra-cluster medium as well as interactions in the extragalactic space. The black points are the upper limit on the photon flux from Perseus Cluster placed by MAGIC. \cite{Aleksic2012_AA541A99}.}
  \label{fig:pedict}
\end{center}
\end{figure}

Propagation in the interstellar medium (ISM) was also modeled to account for particle interactions and energy losses prior to escaping the cluster environment. 
Using the best-fit parameters derived from spectral and compositional data (from \cite{Marafico:2024}), we determined the burst rate per year and normalized the expected neutrino and photon fluxes, which are presented with their respective uncertainties. The photons and neutrino fluxes are shown in figure \ref{fig:pedict}. 
As can be appreciated, the neutrino and gamma‑ray cases differ significantly in both expected flux abundance and characteristic energy ranges.  In the case of neutrino observatories, public instrument response functions (IRFs) are generally provided only for point‑like or narrow extended sources, and Instrument Responde functions tailored to very large angular regions like the Virgo Cluster are not readily available.  This lack of publicly accessible extended‑source IRFs makes it difficult to perform accurate sensitivity estimates without direct collaboration or proprietary Monte Carlo tools.  



For the {neutrino channel}, we note that the flux at Earth cannot be effectively constrained using the point-like sensitivity of Auger, as discussed by the Pierre Auger Collaboration (\cite{Auger2019PointNeutrino}). In particular, the angular size of Virgo, around six degrees, makes the situation worse: the extended source introduces a strong background from atmospheric neutrinos, reducing sensitivity. A more suitable approach would be an extended-source analysis using IceCube or KM3NeT, once public IRFs become available. For the {photon channel}, the situation is somewhat different. The expected photon flux is strongly affected by reprocessing in the source environment and during propagation in extragalactic space. Still, for compact emission regions (within about $0.15^\circ$), upper limits from MAGIC observations \cite{Aleksic2012_AA541A99} already provide useful constraints. In particular, the SMD scenario is disfavored by these limits, while the SFRD scenario remains consistent within one sigma. Future facilities such as CTA, with improved sensitivity, will allow us to tighten these constraints further. Upcoming observatories will improve our ability to test these scenarios, particularly in the context of ultra-high-energy cosmic-ray transients. Nevertheless, the constraints obtained from current UHECR measurements remain robust and provide an essential baseline.  

\section{Constraints on UHECR accelerators}

While the lack of publicly available IRFs for extended‐source neutrino searches and the computational complexity of large‐region gamma‐ray sensitivity calculations precludes us from imposing additional constraints using neutrino and $\gamma$‑ray observations, the limits derived from the UHECR spectrum and mass composition remain robust and continue to delimit the viable source classes.  These constraints, illustrated in Figure~\ref{fig:constraints}, are subdivided into high‑luminosity and low‑luminosity long GRBs (\textit{hl}L‑GRBs and \textit{ll}L‑GRBs, respectively, with the dividing line at $10^{50}\,$erg\,s$^{-1}$), as well as ultra‑long GRBs whose prompt emission persists for several kiloseconds (see references in Table in \cite{Marafico:2024}). 
As shown by the color scale in Figure~\ref{fig:constraints}, non‑jetted TDEs and shock breakout events (SBOs) generally fail to satisfy the Hillas–Lovelace–Waxman–Blandford criterion, and jetted TDEs occur too rarely to account for the observed UHECR flux.  Although the kinetic energies of short GRB ejecta are typically too low, more precise measurements of their outflow dynamics may further clarify their candidacy.  In contrast, all subclasses of long GRBs fulfill the necessary criteria, with \textit{hl}L‑GRBs residing in an especially favorable region of parameter space.  
We conclude that, with the current setup,  the only stellar-sized transients that satisfy both Hillas' and our criteria are long gamma-ray bursts. {In addition, 
We report the estimated local merger rates for binary neutron stars, neutron star–black hole systems, and binary black hole mergers \cite{LVK2023b,AandA2018}. Although the kinetic energy per burst of these candidates is not directly constrained, we note that binary neutron star mergers alone can already satisfy the requirements imposed by UHECR observations.

\begin{figure}[htp]
\centering
\includegraphics[width=0.9\columnwidth]{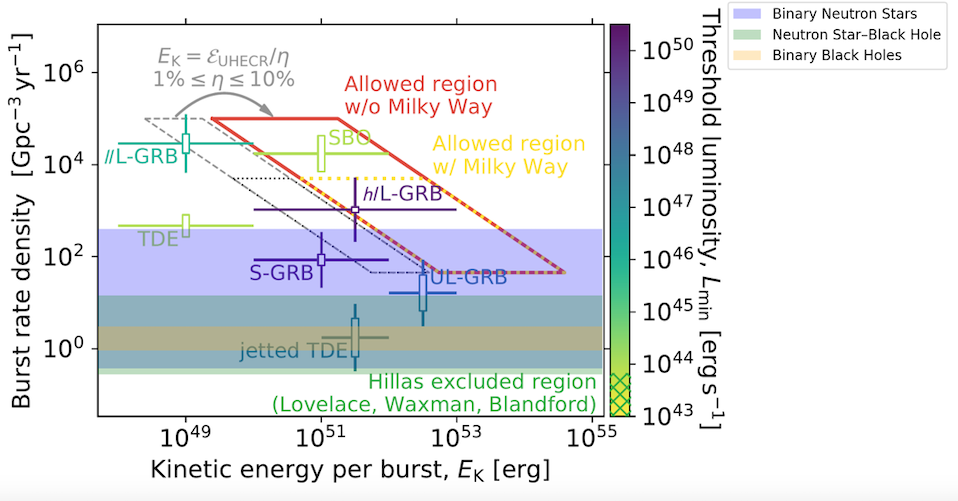}
\caption{UHECR burst rate as a function of kinetic energy of the outflow. The allowed regions are based on an efficiency of conversion of kinetic energy to particles, $\eta$, between $1\,\%$ and $10\,\%$. Rectangular markers indicate the statistical uncertainty associated with the number of X-ray bursts observed for each type of source. The vertical error bars show the range associated with the beaming correction factor, while the horizontal error bars illustrate the range of kinetic energy per burst in each population. The color bar shows the threshold luminosity above which each source density is measured. The low luminosities excluded by the Hillas-Lovelace-Waxman-Blandford criterion are shown as a green hatched region of the color bar. Horizontal shaded bands mark the corresponding range of local merger rates inferred from gravitational-wave observations.}
\label{fig:constraints}
\end{figure}
In the future, several upcoming experiments, including \textit{GRAND}, \textit{POEMMA}, and \textit{GCOS}, are expected to significantly enhance sensitivity to cosmogenic neutrinos and photons. These facilities will be capable of probing the parameter space relevant to Cluster-origin UHECRs and may provide crucial observational evidence of acceleration processes in galaxy clusters. In particular, detection of neutrino fluxes consistent with our predictions would strongly support the transient scenario.
\section{Aknowdlegements}
AC, JB and OD gratefully acknowledge funding from ANR, via the grant MultI-messenger probe of Cosmic Ray Origins (MICRO), ANR-20-CE92-0052.

\end{document}